\documentclass[12pt]{JINST}

\usepackage{graphicx}
\usepackage{amssymb}
\usepackage{epstopdf}

\newcommand{\dedx}{\mbox{${\rm d}E/{\rm d}x$}}

\newcommand{\rphi}{\mbox{$r \! \cdot \! \phi$}}

\title{Comments on TPC and RPC calibrations reported 
by the HARP Collaboration}

\author{V~Ammosov$^3$, 
I~Boyko$^2$, 
G~Chelkov$^2$, 
D~Dedovitch$^2$, 
F~Dydak$^1$, 
A~Elagin$^2$\thanks{Now at Texas A\&M University, 
College Station, USA.},
V~Gapienko$^3$,
M~Gostkin$^2$, 
A~Guskov$^2$,
Z~Kroumchtein$^2$,
V~Koreshev$^3$, 
Yu~Nefedov$^2$,
K~Nikolaev$^2$,
A~Semak$^3$,
Yu~Sviridov$^3$,
E~Usenko$^3$,
J~Wotschack$^{1}$\thanks{Corresponding author; 
E-mail: Joerg.Wotschack@cern.ch},
V~Zaets$^3$ and
A~Zhemchugov$^2$ \\
\llap{$^1$}CERN, Geneva, Switzerland \\
\llap{$^2$}Joint Institute for Nuclear Research, Dubna, Russia \\
\llap{$^3$}Institute for High Energy Physics, Protvino, Russia}

\abstract{The HARP Collaboration recently published calibrations of 
their TPC and RPC detectors, and differential cross-sections of 
large-angle pion production in proton--nucleus collisions. 
We argue that these calibrations are biased and cross-sections 
based on them should not be trusted.}


\begin{document}

\section{Why these `Comments'?}

The larger part of the HARP Collaboration 
(hereafter referred to as `HARP' or `authors') recently published
a paper entitled `Momentum scale in the 
HARP TPC'~\cite{arXivPreprintTPC}. Therein, they claim that they 
calibrated the momentum scale of the HARP TPC
with a precision of 3.5\%. They published a paper
entitled `The time response of glass resistive plate chambers 
to heavily ionizing particles'~\cite{arXivPreprintRPC}. Therein, 
they claim a 500~ps time advance of 
protons with respect to minimum-ionizing 
pions in the HARP multi-gap timing 
RPCs~\cite{HARPTechnicalPaper}--\cite{IEEERebuttal}. Further, they published differential 
cross-sections of pion production on Ta~\cite{OffTaPub}, 
C, Cu and Sn~\cite{OffCCuSnPub}, and 
Be, Al and Pb~\cite{OffBeAlPbPub} targets.

We, also members of the HARP Collaboration and referred to
as `HARP-CDP'\footnote{CDP stands for CERN--Dubna--Protvino}, have not signed the above-cited papers because we are unable to take responsibility for the reported calibrations and physics results. 

We shall argue that there is no reason to invoke a new detector 
physics effect in multi-gap timing resistive plate chambers (RPCs), yet there are good 
reasons why HARP's time projection chamber (TPC) and RPC calibrations should not be trusted,
and also cross-sections of large-angle pion production on nuclear targets based on them.

\section{HARP's biased $p_{\rm T}$ scale and bad $p_{\rm T}$ resolution}
\label{pTdiscussion}

The performance of the HARP TPC was affected by 
dynamic track distortions that were primarily caused by the 
build-up of an Ar$^+$ ion cloud during the 400~ms long spill 
of the CERN Proton Synchrotron. 
This ion cloud emanates from the TPC's sense wires and 
drifts across its active volume toward the high-voltage 
membrane\footnote{The cause of this hardware problem, the physics 
of the track distortions, their quantitative assessment, 
and their corrections, are described in 
Refs.~\cite{distortions}--\cite{distortions4}.}.

These dynamic track distortions 
increase approximately linearly with time in the spill. 
Their size in the 
\rphi\ coordinate typically reaches 15~mm, at small radius, 
at the end of the spill. That exceeds the TPC's design \rphi\ 
resolution of 500~$\mu$m by a factor of 30 and therefore
requires very precise track distortion corrections.
 
The authors published two quite different analysis concepts to 
deal with dynamic track distortions. 

The first concept is to use 
only the first 100 events out of typically 300 events in the whole 
accelerator spill. From the `physics benchmark' of 
proton--proton elastic scattering they claim that dynamic
distortions do not affect the quality of the first 
100 events, and hence dynamic track distortions
need not be corrected at all. 
The second concept is a correction of the distortions based
on a specific radial dependence of the charge density 
of the Ar$^+$ ion cloud.

In the HARP TPC, with a positive magnetic field polarity, 
dynamic distortions shift cluster positions such that 
positive tracks are biased toward higher $p_{\rm T}$ 
(conversely, negative tracks are biased toward 
smaller $p_{\rm T}$). The authors  
chose---in principle correctly---to fit 
TPC tracks with the constraint of the beam point 
because the increased lever arm permits
an approximate doubling of
the $p_{\rm T}$ precision. While the beam point 
remains unaffected, the cluster positions get shifted by
dynamic distortions. 
Assigning a sufficiently small 
position error to the beam point renders its weight 
(the inverse error squared) in the track fit so large that 
positive tracks get biased toward lower $p_{\rm T}$, i.e., the 
trend of the bias is even reversed with respect to the fit 
without beam point. This---artificially enforced---decrease 
of the $p_{\rm T}$ of positive tracks with the time in the spill
is demonstrated in the right panel of figure~15 in Ref.~\cite{OffTaPub}.

This makes clear that the weight assigned in the track fit 
to the beam point is of paramount importance. Despite this
importance, the weight of the beam point has never been
quantitatively stated by HARP. 

Because the bias has different size and opposite sign depending
on whether the beam point has been used in the fit or 
not, we recall   
that in the cross-section results reported by HARP 
the fit with the beam point has been used,
but not in all their `physics benchmarks'.

There is no claim that HARP's $p_{\rm T}$ scale
is wrong {\it per se\/}. Rather, we claim that HARP's initially
(more or less) correct $p_{\rm T}$ scale develops a  
bias that increases about linearly with the time in the spill.
This bias is a direct consequence of the development of
dynamic track distortions with time in the spill. This means 
that the percentage of the claimed bias is not constant 
but proportional to $p_{\rm T}$. This means that the 
claimed bias is {\it a priori\/} different from 
data set to data set since dynamic distortions are
different in different data sets. Therefore, 
conclusions on a bias in one data set 
(e.g., elastic scattering of 3~GeV/{\it c} 
protons on protons at rest) cannot be applied 
quantitatively to other data sets. 

In their first analysis concept (that underlies the cross-sections
published in Refs.~\cite{OffTaPub}--\cite{OffBeAlPbPub}),
the authors fit the distorted track together with the undistorted
beam point. The beam point is assigned a weight {\it `similar
to a TPC hit'}~\cite{OffTaPub} which 
implies that the beam point's error is constant and not what it 
must be: the convolution of the errors 
of two extrapolations to the interaction vertex, of the 
beam particle's trajectory and of the secondary track's trajectory. 
Primarily because of the momentum-dependence of multiple 
scattering, the correct error of the beam point 
varies considerably for different
beam momenta and from track to track. The authors
fit a circle to distorted TPC cluster positions that 
deviate in a radius-dependent way by up to 5~mm from their 
nominal positions, and to the undistorted beam point that has 
a wrong weight in the fit. Under such circumstances, the fit of 
$p_{\rm T}$ cannot be unbiased.

How large is the bias in this concept? The authors give the answer 
themselves in the upper left panel of figure~17 in 
Ref.~\cite{OffTaPub} where they show the measurement of the
specific ionization \dedx\ of protons as a function of 
momentum. One reads off that an 800~MeV/{\it c} proton is 
measured with a momentum of 650~MeV/{\it c}. 
From this $\sim$20\% scale error for positive particles
at $p = 800$~MeV/{\it c}, one infers a scale
error of $\sim$20\% in the opposite direction for negative 
particles. Expressed as a shift of $q/p_{\rm T}$ (where $q$ denotes the
particle's charge), the bias is of order
$\Delta (q/p_{\rm T}) \sim +0.3$~(GeV/{\it c})$^{-1}$ for
positive magnet polarity. 

The effect of this bias is well visible in a comparison 
of HARP's $q/p_{\rm T}$ spectrum with the one from our 
group, see figure~10 in Ref.~\cite{FinalReportOfRBH}.

In their second analysis concept, the authors apply a correction of
dynamic track distortions and use data from the whole spill.
The correction stems from the electric field of a 
charge density of Ar$^+$ ions that falls with the radial 
distance $R$ from the beam like $1/R^2$~\cite{uvpageSPSC}.

Is a $1/R^2$ distribution realistic?      
The answer is no. The radial
charge distribution depends on beam energy, beam polarity, 
beam intensity, beam scraping, target type, photon conversion 
in materials and spiralling low-momentum electrons. Therefore,
the correction algorithm cannot be expected to work with
adequate precision. 

\begin{figure}[ht]
\begin{center}
\vspace*{3mm}
\includegraphics[width=10cm,]{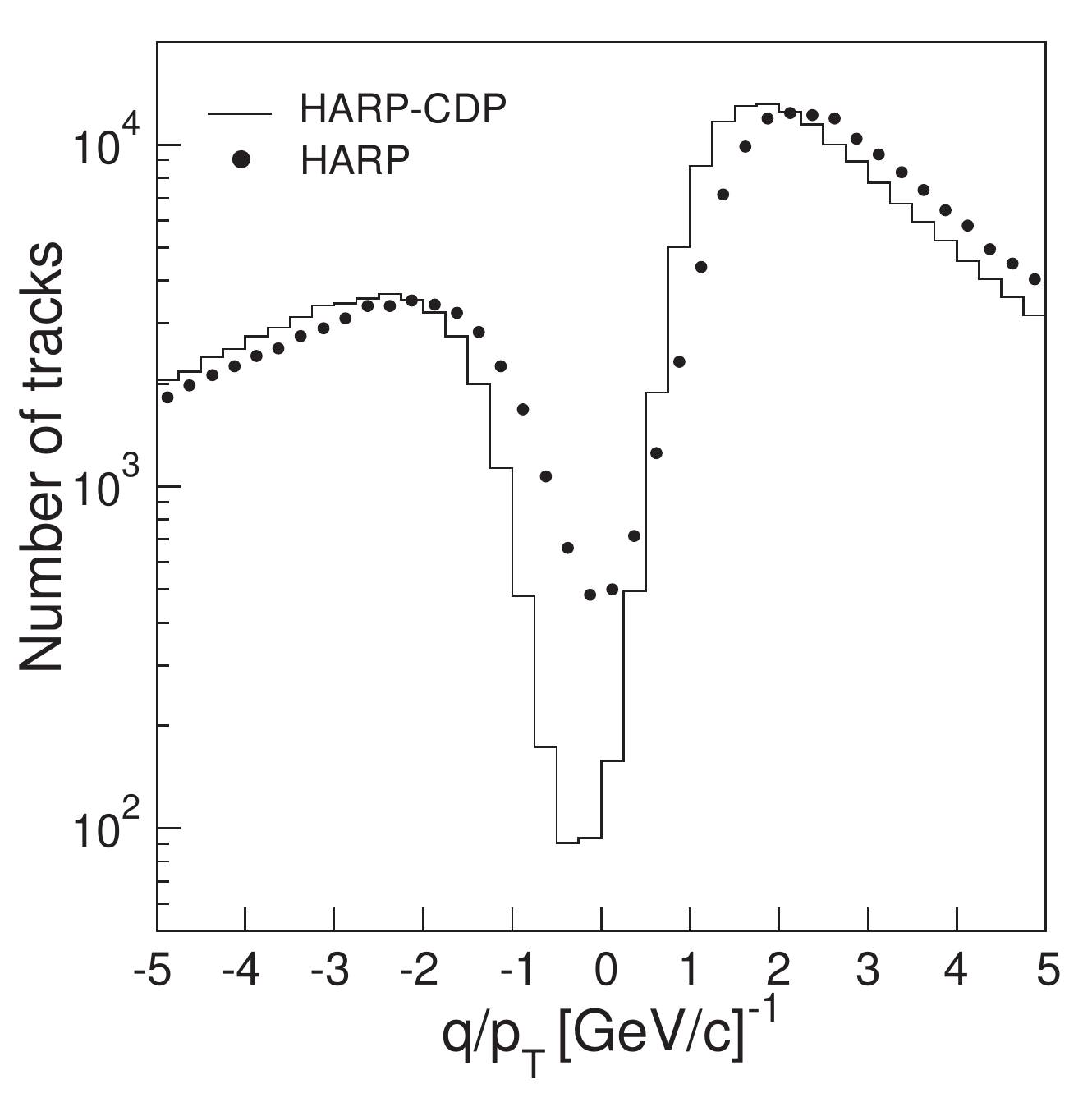}
\caption{$q/p_{\rm T}$ spectra of secondary particles from the
interactions of $+8.9$~GeV/{\it c} protons in a 5\% $\lambda_{\rm abs}$
Be target; the black points are taken from figure~14 in 
Ref.~\cite{arXivPreprintTPC}, the histogram represents the
HARP--CDP analysis of the same data.}
\label{spectra}
\end{center}
\end{figure}

This expectation is confirmed by   
the difference between the data shown in figure~14 in 
Ref.~\cite{arXivPreprintTPC} and the same data analysed
by our group, see figure~\ref{spectra} that shows the 
$q/p_{\rm T}$ spectra of secondary particles from the
interactions of $+8.9$~GeV/{\it c} protons in a 
5\% $\lambda_{\rm abs}$ Be target.

The difference of the spectra is again 
consistent with a HARP bias of  
$\Delta (q/p_{\rm T}) \sim +0.3$~(GeV/{\it c})$^{-1}$
with respect to our results from the same 
data.

The authors claim~\cite{HARPTechnicalPaper,OffRPCPaper,OffTaPub} 
a resolution of
\begin{displaymath}
\sigma (p_{\rm T})/p_{\rm T} = 
  (0.25 \pm 0.01)p_{\rm T} + (0.04 \pm 0.005) \; ({\rm GeV}/c)^{-1} \; 
\label{offpTresolution}
\end{displaymath}
or, approximately, $\sigma (1/p_{\rm T}) \sim 0.30$~(GeV/{\it c})$^{-1}$.  
This claimed resolution refers to fits with 
the beam point included (in fits without the beam point the 
resolution is around 0.60).

The information given by the authors on the experimental $p_{\rm T}$ resolution
for fits with the beam point included (on which all reported
cross-sections are based) is very scarce. It consists of a mere 
three points in figure~9 in Ref.~\cite{OffTaPub}.
One reads off the resolution 
$\sigma (1/p_{\rm T}) \sim 0.5$~(GeV/{\it c})$^{-1}$. 
Although this resolution represents a convolution with the
\dedx\ resolution, it is hardly compatible 
with the claimed 0.30~(GeV/{\it c})$^{-1}$. 

Confirmation that the $p_{\rm T}$ resolution 
is much worse than claimed is given in 
Refs.~\cite{OffRPCPaper} and \cite{IEEERebuttal}. Therein, the  
RPC time-of-flight resolution of $p \sim 200$~MeV/{\it c} pions
that is equivalent to the $p_{\rm T}$ resolution in the
TPC is quoted as 260~ps. As succinctly proven in 
Refs.~\cite{IEEEComments} and \cite{WhiteBookAddendum2}, 
a time-of-flight resolution of 
260~ps of pions with $p_{\rm T} = 200$~MeV/{\it c} is
equivalent to a resolution $\Delta p_{\rm T}/p_{\rm T}$ of 46\%, 
which is worse by a stunning factor of 4.6 than the 
claimed resolution\footnote{This result is obtained when 
taking literally two more claims by HARP: a beam-particle
timing resolution of 70~ps and an RPC timing resolution of 141~ps;
however, it is more likely that the overall discrepancy of
4.6 stems from all three sources and not only from the
bad $p_{\rm T}$ resolution.}.     

Figure~\ref{spectra} also proves that HARP's $p_{\rm T}$ resolution is
much worse than claimed. The depth of the dip at $q/p_{\rm T} = 0$
reflects directly the $p_{\rm T}$ resolution, and HARP's dip is
considerably more shallow than ours. 

The difference between HARP's and our $q/p_{\rm T}$ spectra is  
consistent with a HARP bias of  
$\Delta (q/p_{\rm T}) \sim +0.3$~(GeV/{\it c})$^{-1}$,
and a HARP resolution of 
$\Delta (q/p_{\rm T}) \sim 0.55$~(GeV/{\it c})$^{-1}$.

The discrepancy between the $q/p_{\rm T}$ spectra means
that cross-sections are different by factors of up to two.

The authors claim that results from the second concept 
of correcting dynamic
track distortions and using the data from the full spill,
is in {\it `excellent agreement'}~\cite{arXivPreprintTPC}
with results from the first concept of not
correcting for dynamic track distortions and using the
first 30\% of the spill only. 

We agree that there is no difference in the 
results from these two concepts.
Both are affected by a comparable 
$p_{\rm T}$ bias and a comparably bad $p_{\rm T}$ resolution.
That the biases in HARP's two 
analysis concepts happen to have the same 
size and sign, is accidental.

\section{HARP's `500~ps effect'}

The authors reported in Ref.~\cite{HARPTechnicalPaper} 
a 500~ps advance of the RPC timing signal of protons with respect to
the one of pions. They confirmed their discovery  
in three subsequent publications~\cite{NIMRebuttal}--\cite{IEEERebuttal}, and most recently in Ref.~\cite{arXivPreprintRPC}.
In the latter paper, the authors acknowledge that
{\it `...it has been pointed out that a similar behaviour
can be obtained when a systematic shift in the measurement of
momentum is present'\/} but conclude that
{\it `Momentum measurement biases in the TPC, if any, have been
eliminated as possible cause of the effect.'}

In stark contrast, our group's interpretation of the authors' result 
is that their $p_{\rm T}$ scale is systematically 
biased by $\Delta(1/p_{\rm T}) \sim 0.3$~(GeV/{\it c})$^{-1}$ which
leads to the prediction of a longer time of flight for non-relativistic
protons (whereas the time of flight of relativistic pions is unchanged).
In turn, if the proton momentum is considered correct, the
RPC timing of protons would appear to be advanced. 

The relevant experimental variable is the proton
time of flight as measured by the RPCs minus the time of flight
calculated from the proton momentum.

\begin{figure}[ht]
\begin{center}
\vspace*{3mm}
\includegraphics[width=10cm,]{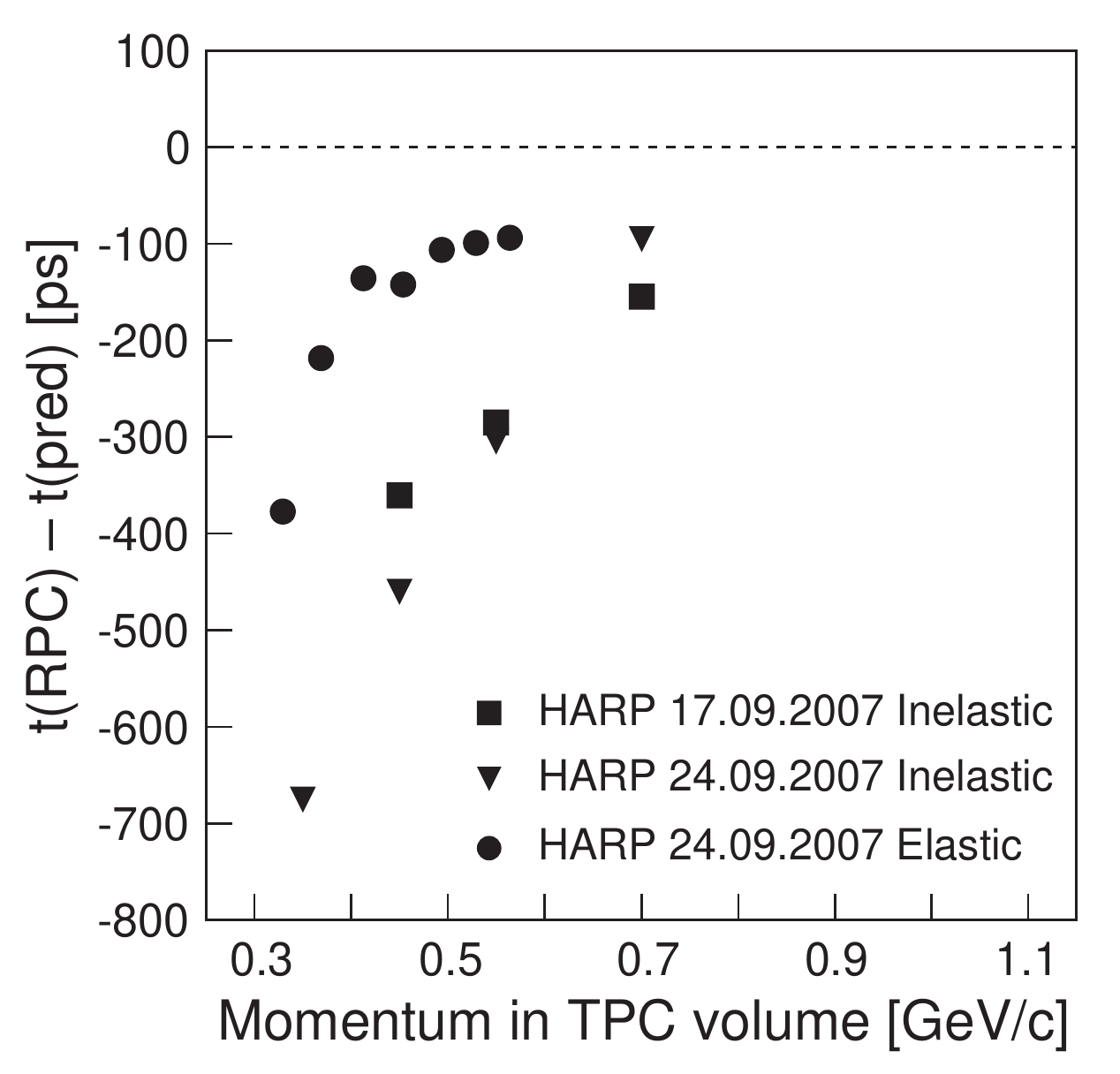}
\caption{Time advance of protons (HARP data): time of flight
measured by the RPCs minus the time of flight calculated from the
momentum measured in the TPC (`Inelastic') and calculated from the
kinematics of proton--proton elastic scattering 
(`Elastic'), respectively.}
\label{panmaneff_oh}
\end{center}
\end{figure}

Figure~\ref{panmaneff_oh} shows HARP's respective data, taken from
their most recent papers~\cite{arXivPreprintTPC} (17 Sep 2007) and 
\cite{arXivPreprintRPC} (24 Sep 2007), data which are based on
their $p_{\rm T}$ measurement in the TPC and hence affected
by a bias in the TPC $p_{\rm T}$ scale\footnote{All data shown in this Section
refer to the RPC padring 3, i.e., to tracks with polar angles $\Theta \sim$ 55--80$^\circ$.}. Also shown are
data from the calculated momentum of recoil protons in 
elastic proton--proton scattering,
published by HARP in Ref.~\cite{arXivPreprintRPC},
data that are not affected by a bias in the $p_{\rm T}$ 
measurement in the TPC. 

All three data sets should show the same time advance 
but disagree seriously with each other. This hardly supports  
the notion of a novel detector physics effect. 

\begin{figure}[ht]
\begin{center}
\vspace*{3mm}
\includegraphics[width=10cm,]{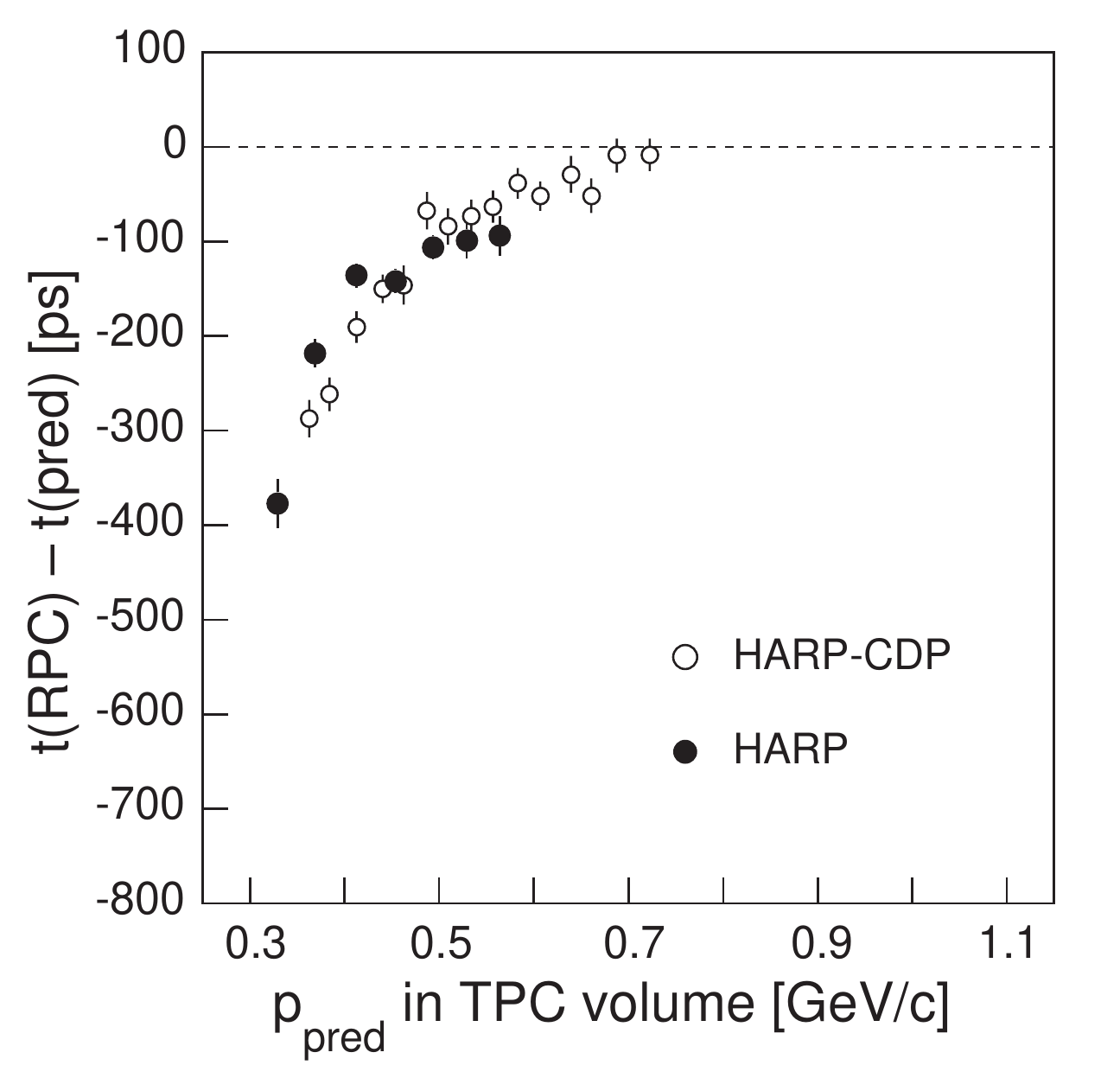}
\caption{Time advance of protons (HARP data and HARP--CDP data): time of flight
measured by the RPCs minus the time of flight calculated from the
kinematics of proton--proton elastic scattering.}
\label{panmaneff_h2}
\end{center}
\end{figure}

Figure~\ref{panmaneff_h2} shows the comparison of HARP and HARP--CDP data
on the timing difference of recoil protons from elastic 
proton--proton scattering. 
There is good agreement between the data which confirms that
both HARP and HARP--CDP correctly calibrated the RPCs 
with relativistic pions. The data from elastic proton--proton 
scattering are consistent with the theoretically expected 
time advance (for the calculation of the theoretically expected
time advance, we refer to our pertinent 
discussion in Ref.~\cite{RPCpub}).

\begin{figure}[ht]
\begin{center}
\vspace*{3mm}
\includegraphics[width=10cm]{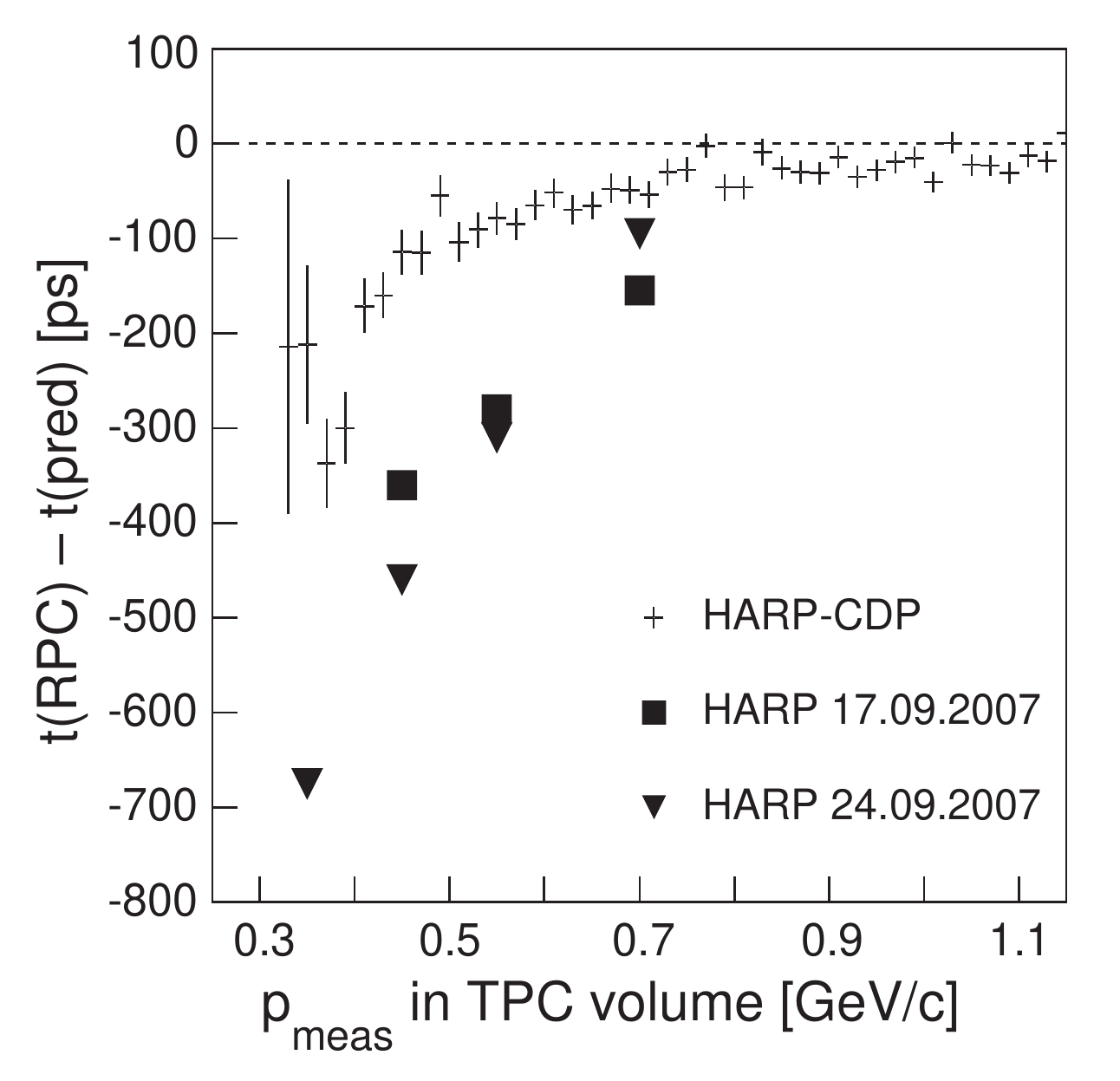}
\caption{Time advance of protons (HARP data and HARP--CDP data): time of flight
measured by the RPCs minus the time of flight calculated from the
momentum measured in the TPC.}
\label{panmaneff_be}
\end{center}
\end{figure}

Figure~\ref{panmaneff_be} shows the comparison of HARP and HARP--CDP
data for the case that the $p_{\rm T}$ reconstruction in the
TPC is used to determine the time of flight of the recoil proton.
While the HARP--CDP data confirm the results from proton--proton
elastic scattering, the HARP data are inconsistent with
these results. 

\begin{figure}[ht]
\begin{center}
\vspace*{3mm}
\includegraphics[width=10cm]{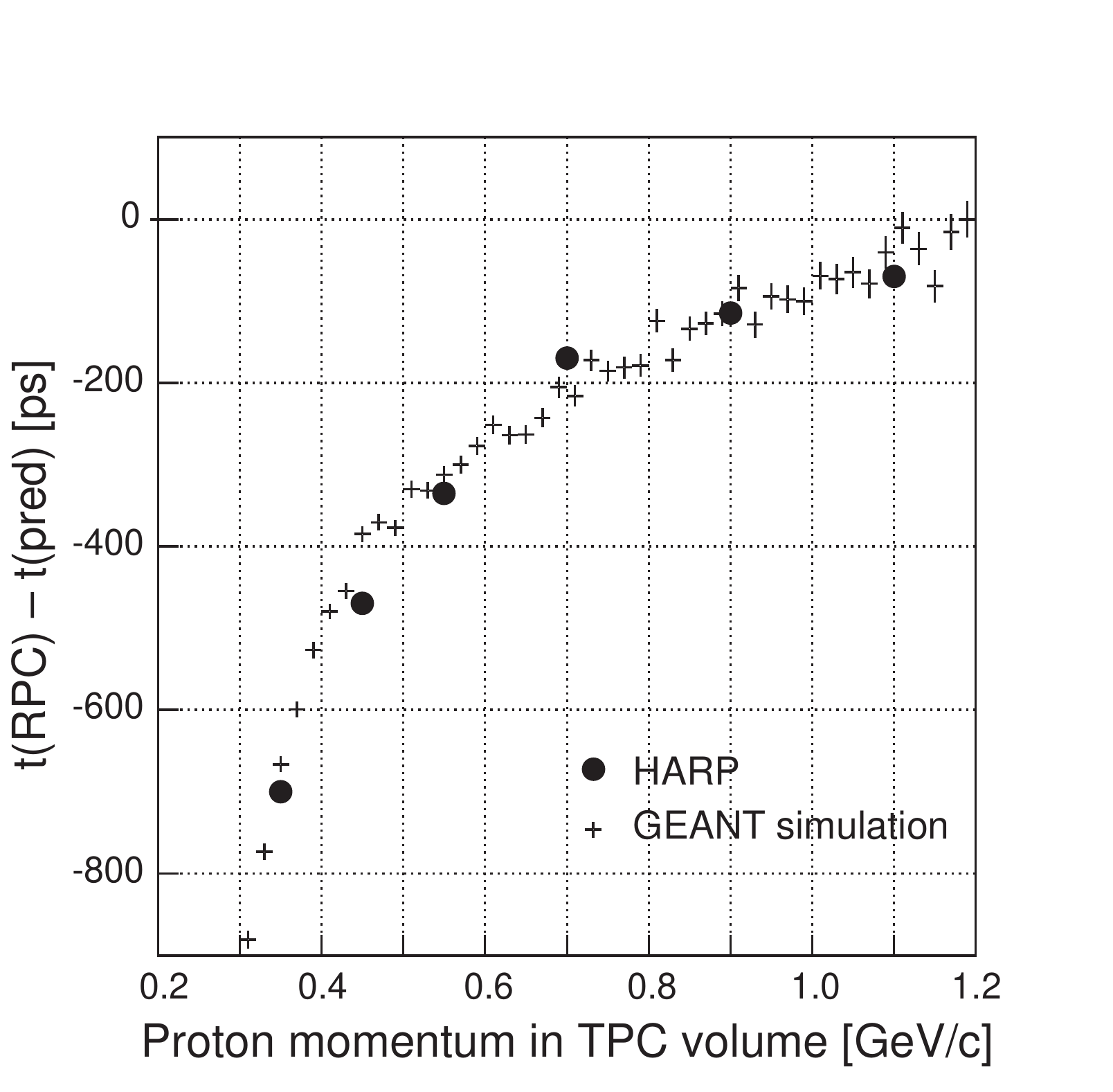}
\caption{HARP's time advance of protons (black points; data from
Ref.~\cite{arXivPreprintRPC}) compared with
a simulation of the time advance that will result from a bias 
$\Delta(1/p_{\rm T}) \sim 0.30$~(GeV/{\it c})$^{-1}$.}
\label{explainpanmaneff}
\end{center}
\end{figure}

Figure~\ref{explainpanmaneff} shows that HARP's time advance 
of protons (black points; data from Ref.~\cite{arXivPreprintRPC}) 
is satisfactorily explained by
a simulation of the time advance that results from a bias 
$\Delta(1/p_{\rm T}) \sim 0.30$~(GeV/{\it c})$^{-1}$. 

There is no need and no room for a novel detector physics effect. 

\section{HARP's `physics benchmark'}

The authors make extensive use of elastic scattering of 
3 and 5~GeV/{\it c} protons and pions on protons at rest 
to support the claim that their $p_{\rm T}$ scale is 
correct within 3.5\%. In the following we show that 
their arguments are not conclusive.

\subsection{Fits of recoil protons with and without beam point}

\begin{figure}[htp]
\begin{center}
\vspace*{3mm}
\begin{tabular}{cc}
\includegraphics[height=8cm]{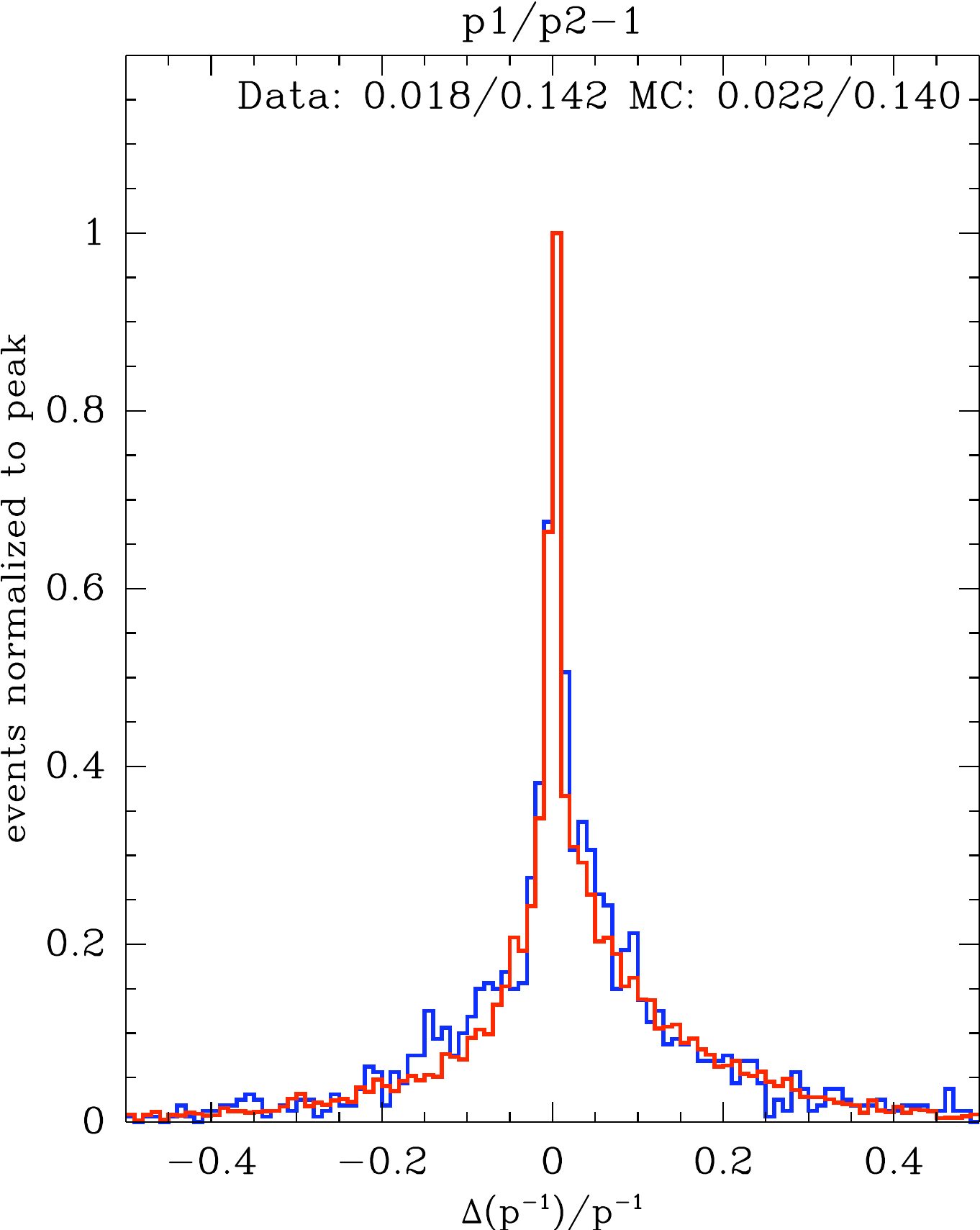} &
\includegraphics[height=8cm]{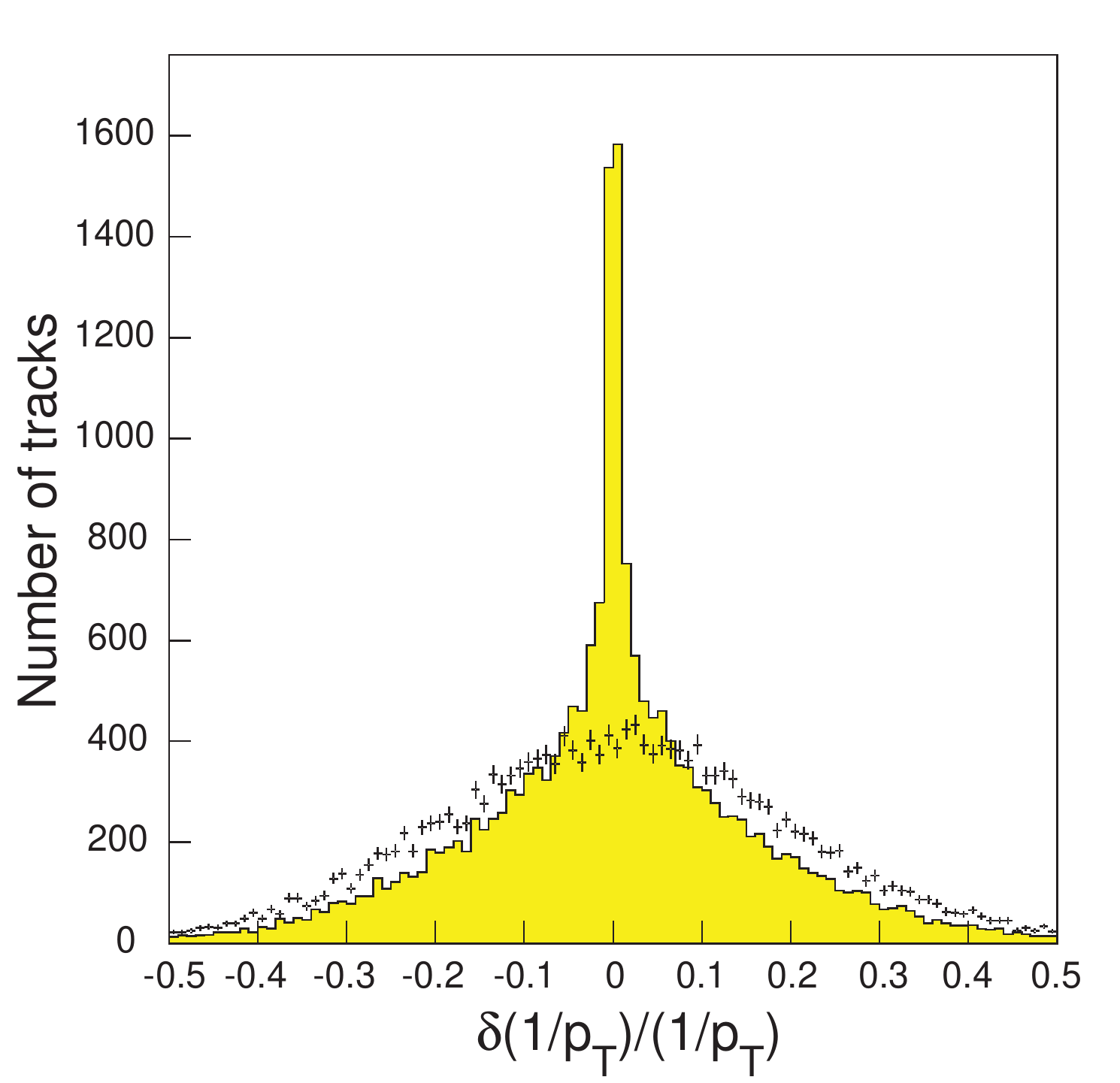} \\
\end{tabular}
\caption{Left: HARP's comparison $(p1-p2)/p2$ of the fit without beam 
point ($p1$) and with the beam point ($p2$), for data and 
Monte Carlo; this figure is a copy from 
figure~4 in Ref.~\cite{arXivPreprintTPC}; right: 
simulation of the expected Gaussian distribution of $(p1-p2)/p2$ 
(crosses) and of its deformation (shaded histogram)  
by HARP's $\cos 2\phi$ factor applied to 
the position error of TPC clusters.} 
\label{OfficialMomComp}
\end{center}
\end{figure}

In stark contrast with our claim of a positive bias in $q/p_{\rm T}$
in fits with the beam point, and a negative bias in 
fits without the beam point, the authors write  
{\it `The ratio of the unconstrained and constrained fits was
checked to be unity with a high precision'} and show figure~4
in Ref.~\cite{arXivPreprintTPC} in support of this claim. 
For its importance, this figure is reproduced in the left panel
of our figure~\ref{OfficialMomComp}. 

One would expect to see a Gaussian distribution in the authors'
variable $(p1-p2)/p2$ ($p1$ is the momentum from a fit without 
the beam point, and $p2$ the momentum from a fit with the 
beam point). Since the claimed resolution with the beam point 
included is 0.30, and without the beam point about 0.60, 
the Gaussian should have a $\sigma \sim 0.50$.  
Their plot shows something very different, though: a narrow spike 
centred at zero, on top of a broad distribution. The authors 
interpret this as evidence that the two fits give the same result. 
 
The spike at zero is an artefact which stems from the 
assignment of a wrong error in the \rphi\ position of clusters:
the authors multiply the \rphi\ error of each TPC cluster 
with $\cos 2\phi$ (a conceptual mistake of their algorithm
as discussed in Ref.~\cite{WhiteBook}) 
and hence produce nearly infinite weights of clusters close 
to the $\phi$ angles 45$^\circ$, 135$^\circ$, 225$^\circ$ 
and 315$^\circ$. In comparison with these wrong large weights, 
the weight of the beam point becomes negligible, which
explains that the fits of tracks close to the singular $\phi$ 
angles yield the same $p_{\rm T}$ with and without 
the beam point. 

The expected Gaussian distribution with $\sigma$ of about 0.50
is indeed visible in their plot: it is the broad distribution below
the artifical spike. This is evident from the right panel 
in figure~\ref{OfficialMomComp}
which shows a simulation how the Gaussian becomes deformed
by the $\cos 2\phi$ term. 

We conclude that the authors did not prove that the fits 
with and without beam point give the same result. Rather, 
they proved that their track fit is seriously compromised.

\subsection{Missing mass from elastic scattering}

The authors write {\it `A fit to the distribution 
[of missing mass squared] provides $M^2_{\rm x} = 
0.8809 \pm 0.0025$~(GeV/{\it c}$^2$)$^2$ in agreement with
the PDG value of 0.88035~(GeV/{\it c}$^2$)$^2$ ... a momentum
scale bias of 15\% would produce a displacement of about
0.085~(GeV/{\it c}$^2$)$^2$ on $M^2_{\rm x}$. As a result, we can
conclude that the momentum scale bias (if any) is
significantly less than 15\%.'}

\begin{figure}[htp]
\begin{center}
\vspace*{3mm}
\includegraphics[width=8cm,angle=-90]{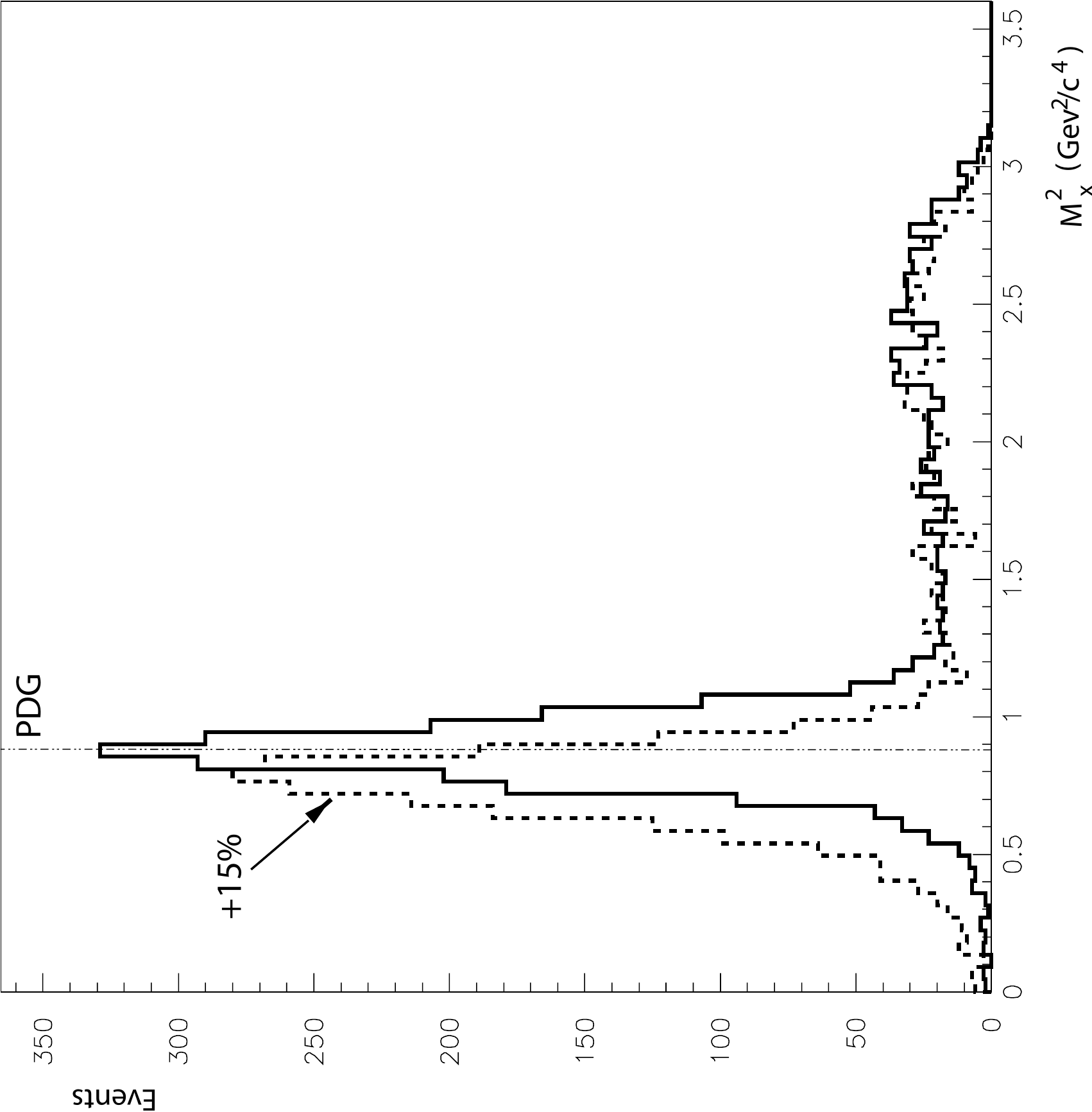}
\caption{HARP's missing mass squared in the elastic scattering of
3~GeV/{\it c} protons on protons at rest; this figure is a copy 
of figure~2 in Ref.~\cite{arXivPreprintTPC}.}
\label{OfficialMissingMass}
\end{center}
\end{figure}

For its importance, their supporting figure~2 in 
Ref.~\cite{arXivPreprintTPC} is reproduced in 
our figure~\ref{OfficialMissingMass}. 

The authors state that the fit of the recoil protons included the
beam point. But they do not give important information: which fraction 
of the spill was used\footnote{We assume that the first 100 
events of the spill were used.}, and they
do not state how the significant energy loss of protons in
materials before the TPC volume was handled\footnote{We assume 
that the proton energy loss was corrected as a function of the 
proton momentum measured in the TPC.}.

Since the beam point was used, the bias in $1/p_{\rm T}$ will
be positive. For the typical $p_{\rm T}$ of the recoil proton
of 0.45~GeV/{\it c}, we estimate from the strength of the 
dynamic distortions in the respective data taking a bias 
$\Delta (1/p_{\rm T}) \sim +0.20$ or, equivalently, 
$\Delta p_{\rm T}/p_{\rm T} \sim -10$\%.
The difference to +15\% in figure~\ref{OfficialMissingMass} 
is important since the missing mass squared is not 
Gaussian-distributed. 

\begin{figure}[htp]
\begin{center}
\vspace*{3mm}
\begin{tabular}{cc}
\includegraphics[width=8cm]{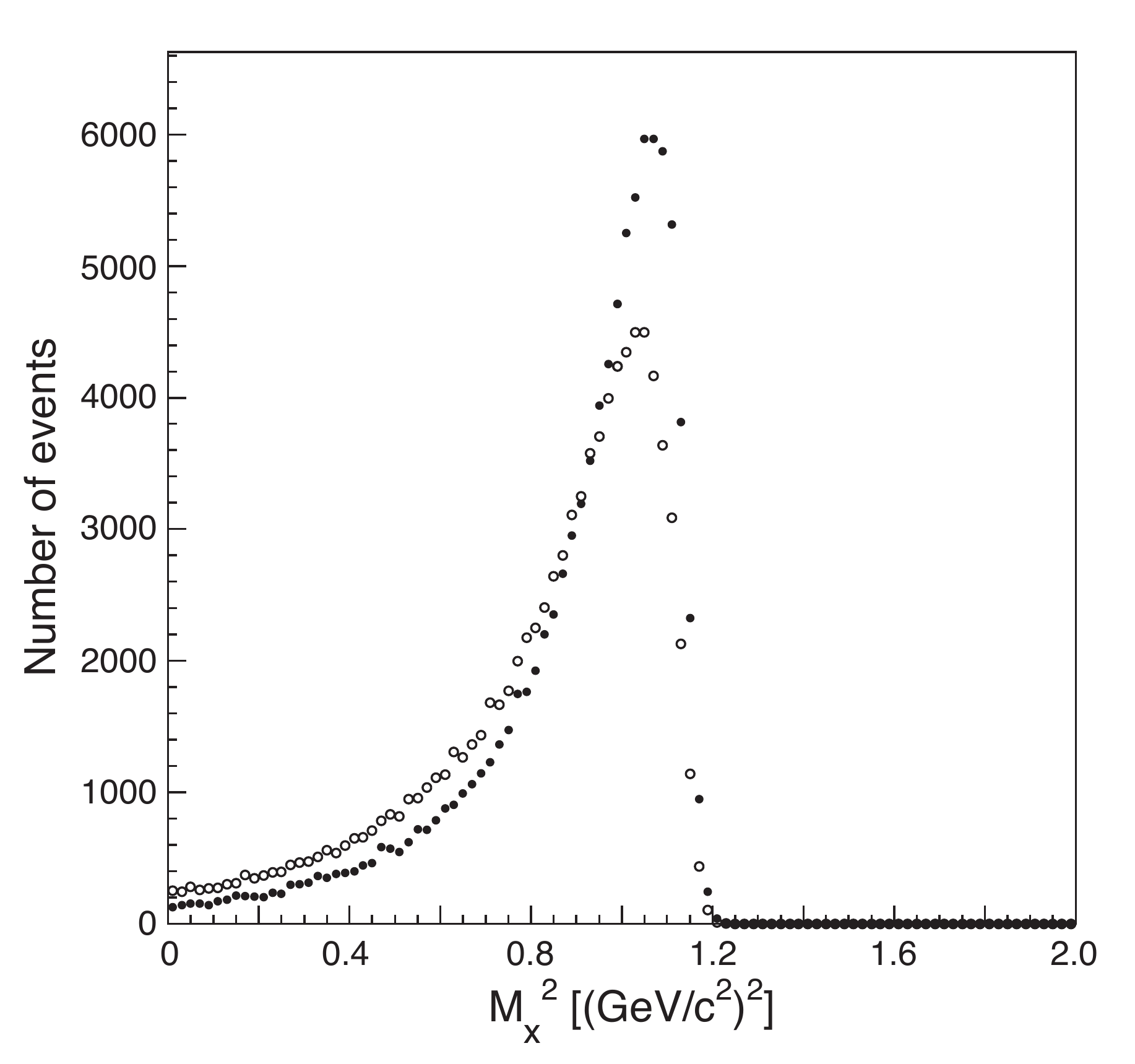} &
\includegraphics[width=8cm]{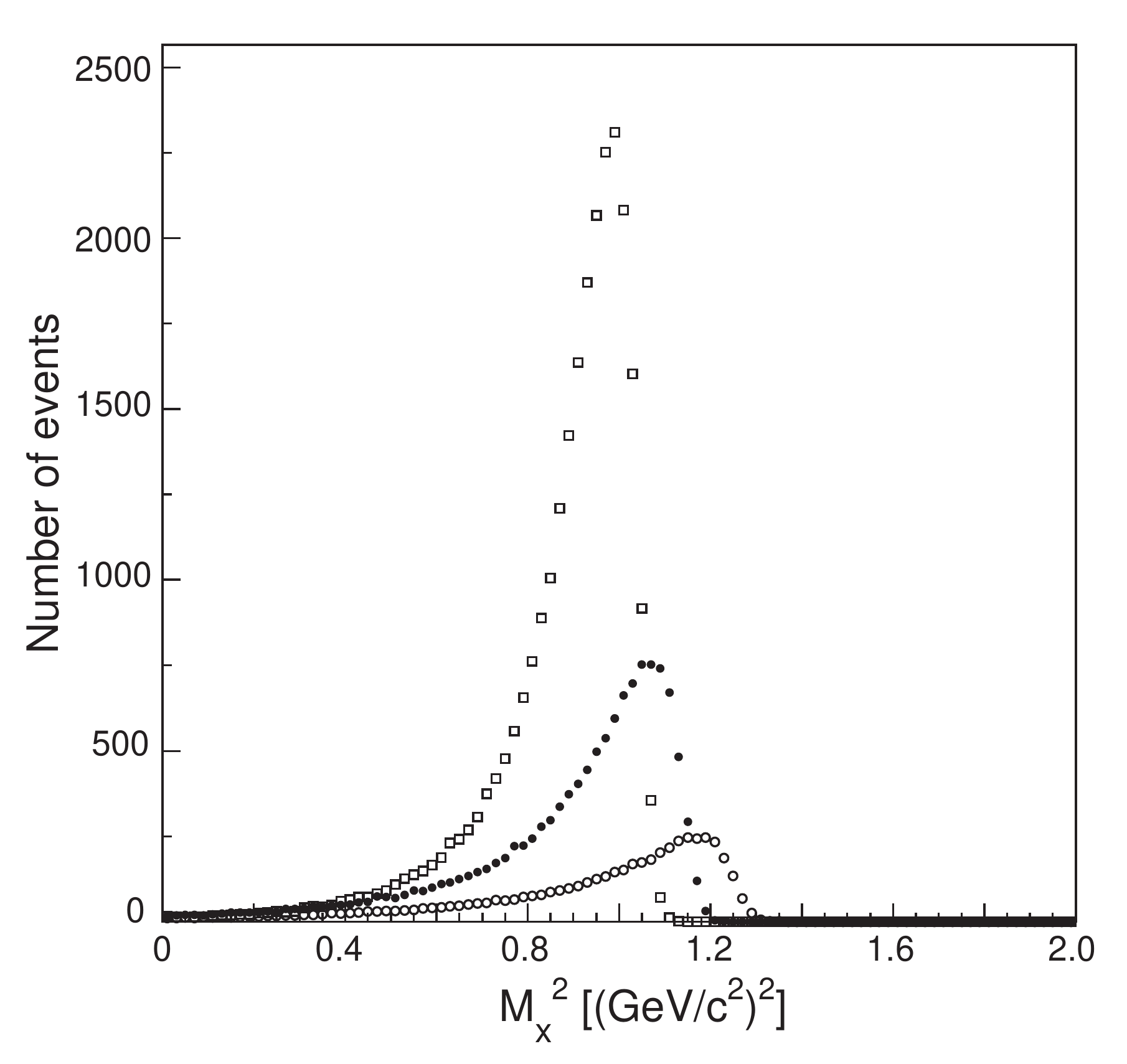} \\
\end{tabular}
\caption{Simulation of the missing mass squared in the 
elastic scattering of 3~GeV/{\it c} protons on protons 
at rest. Left: Simulation for a recoil proton angle of
69$^\circ$ with a resolution of
$1/p_{\rm T}$ of 0.55 and no bias (open circles), with a 
resolution of 0.55 and a bias of +0.20 (black circles).
Right: Simulation with a resolution of $1/p_{\rm T}$ of 0.55 and
a bias of +0.20 for proton recoil angle of 65$^\circ$ (open circles),
69$^\circ$ (full circles) and 73$^\circ$ (open squares).}
\label{MissingMassSimulation}
\end{center}
\end{figure}

Figure~\ref{MissingMassSimulation} shows 
simulations of the missing mass squared in the 
elastic scattering of 3~GeV/{\it c} protons on protons 
at rest. 

The left panel shows the difference, for a proton recoil
angle of 69$^\circ$, between
a distribution with a resolution of
$1/p_{\rm T}$ of 0.55 and no bias, and a distribution with 
the same resolution and a bias of +0.20. The missing mass 
squared distribution is less sensitive
to a $p_{\rm T}$ bias than purported by the authors.

The right panel shows for a resolution of $1/p_{\rm T}$ of 0.55, and
a bias of +0.20, the differences between the proton recoil 
angles of 65$^\circ$, 69$^\circ$ and 73$^\circ$, where the
contributions from the three angles are weighted with their
cross-sections. The sum of the three contribution
may look `Gaussian' but the central value of this `Gaussian' 
cannot be taken as the physical missing mass squared.

The rather erratic nature of results from this analysis
is corroborated by the fit results of the missing-mass-squared
distribution published by the authors in figure~15 in
Ref.~\cite{HARPTechnicalPaper} and reported in 
Ref.~\cite{BorghiThesis}. The result is 15.6$\sigma$ away 
from the PDG value. 

We conclude that the authors did not prove that their
$p_{\rm T}$ from fits with the beam point included 
is unbiased, certainly not with the precision claimed 
by them. Rather, they proved that their analysis
of missing-mass-squared distributions is too simplistic.

\begin{figure}[htp]
\begin{center}
\vspace*{3mm}
\begin{tabular}{cc}
\includegraphics[width=8cm]{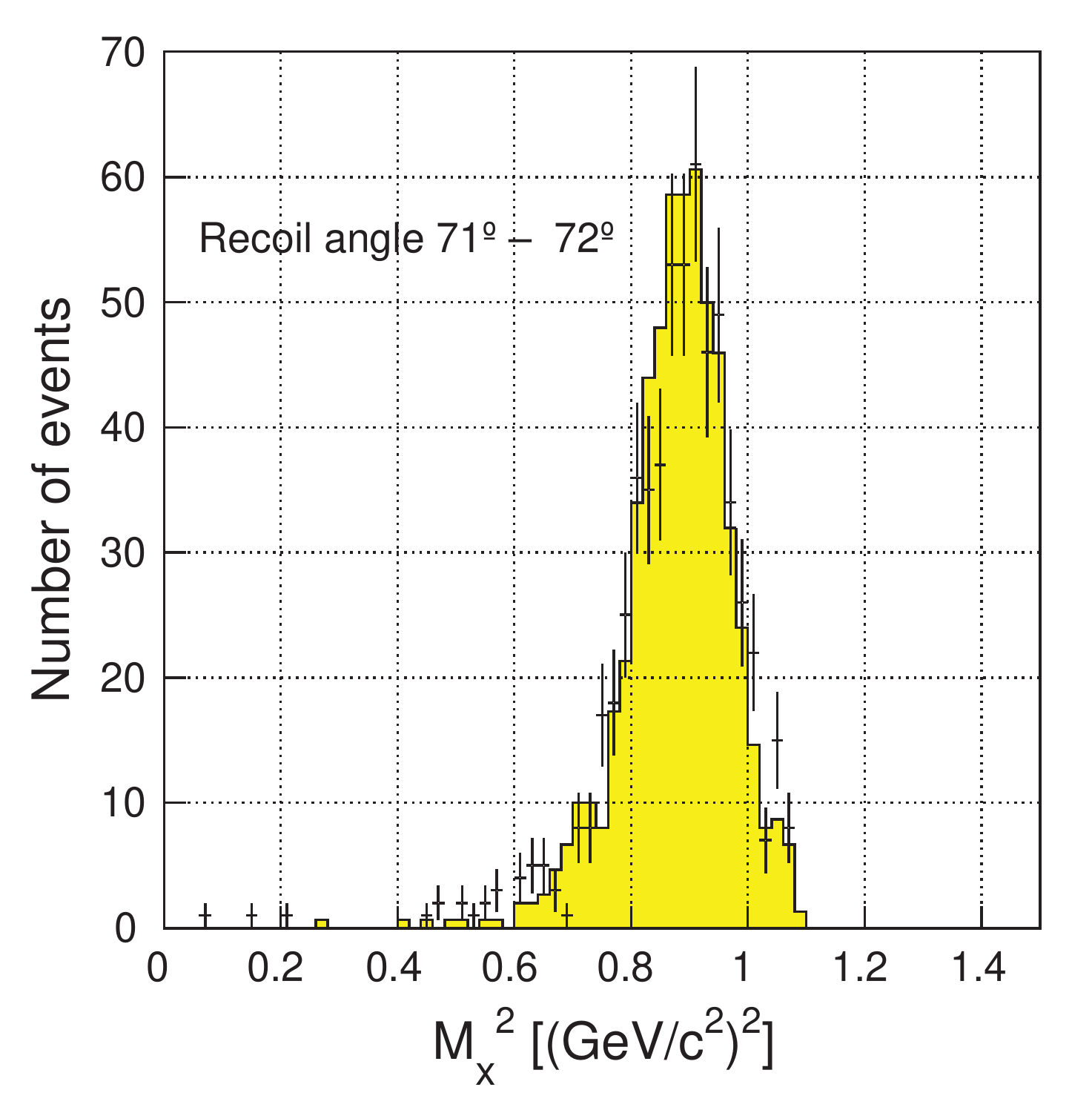} &
\includegraphics[width=8cm]{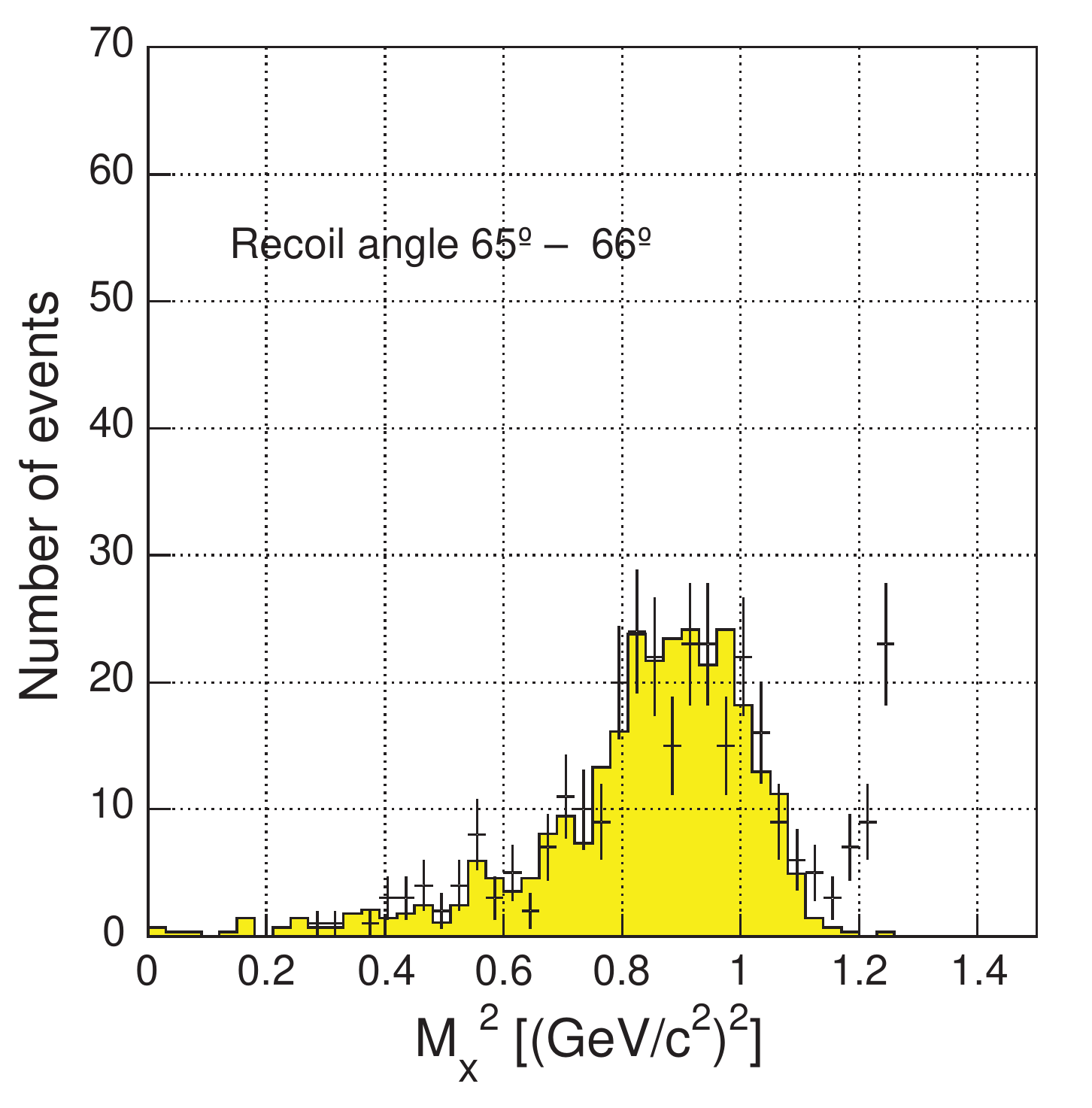} \\
\end{tabular}
\caption{HARP--CDP data on, and GEANT simulation of, 
the missing mass squared in the 
elastic scattering of 3~GeV/{\it c} protons on protons 
at rest; data are shown as shaded histograms, the GEANT simulation
(elastic scattering events only) as crosses; left: proton recoil 
angle between 71$^\circ$ and 72$^\circ$; 
right: proton recoil angle between 65$^\circ$ and 66$^\circ$.}
\label{MissingMassData}
\end{center}
\end{figure}

For comparison, we show in figure~\ref{MissingMassData} our own 
results for the missing mass squared in the elastic scattering of
3~GeV/{\it c} protons on protons at rest, and compare them
with a GEANT simulation. We show the data for two bins in
the proton recoil angle, with a view to highlighting the 
differences both in shape and in rate.  

\subsection{$p_{\rm T}$ scale from elastic scattering}

From the comparison of the momentum  
of recoil protons from the scattering
of 3 and 5~GeV/{\it c} protons and pions on protons at rest
as measured in the TPC, and as predicted from the
measurement of the scattering angle of the forward-going beam particle
in the forward spectrometer, the authors 
conclude that {\it `... a 10\%
bias [of the momentum scale] is excluded 
at 18 $\sigma$ level (statistics only)...'}

In this comparison, a fit without 
the beam point was used. This is important:
(i) the $p_{\rm T}$ resolution will be
about twice worse than in fits with the beam point;
and (ii) the expected bias from dynamic distortions will
have different magnitude and opposite sign compared to the 
bias from fits with the beam point.

Since all data published by HARP are based on fits with 
the beam point, evidence on a bias from dynamic distortions
from fits without beam point is irrelevant; furthermore,
conclusions from the dynamic distortions in one data set 
cannot be applied to another data set. 

We conclude that the authors have not proven that the
$p_{\rm T}$ scale of fits with the beam point is unbiased,
and we could stop our argumentation here.

Nevertheless, we follow the argumentation 
of the authors a bit further.

We note that the authors chose to use only the
first 50 events in the spill which reduces 
the expected bias from dynamic distortions by a factor
of about two compared to the use of the first 100
events in the spill.

We note that for reasons of acceptance, the use of the 
scattering angle of the forward-going beam particle
restricts the recoil protons to the two horizontal
sectors 2 and 5 of the TPC. These are the two sectors
which our group decided not to use for data analysis, for
the much stronger electronics cross-talk and the
many more bad electronics channels in comparison with
the four other TPC sectors, and for the absence
of cross-calibration of performance with cosmic-muon 
tracks.

Still, one is puzzled why HARP find good
agreement between the measured and the predicted
momentum of the recoil proton.

We know from 
our own analysis of the same data that they are
affected by fairly strong dynamic distortions,
albeit smaller in amplitude than  
the $+8.9$ GeV/{\it c} 5\% $\lambda_{\rm abs}$ data shown
in Section~\ref{pTdiscussion},
and with a steeper radial decrease of the
Ar$^+$ ion cloud in the TPC. We have shown in 
Ref.~\cite{TPCpub} that at the start of the spill,
the so-called `margaritka' effect is dominant with a sign
that is opposite to the sign of the so-called 
`stalactite' effect that becomes by far dominant 
later in the spill. Near the start of the spill, there is
a partial cancellation between the two effects (the
cancellation is not complete since the radial distributions
of these track distortions are different). It is
this accidental cancellation that has been 
exploited by HARP to claim that their analysis
is not affected by a bias in the $p_{\rm T}$ 
scale. 

\begin{figure}[htp]
\begin{center}
\vspace*{3mm}
\includegraphics[width=10cm]{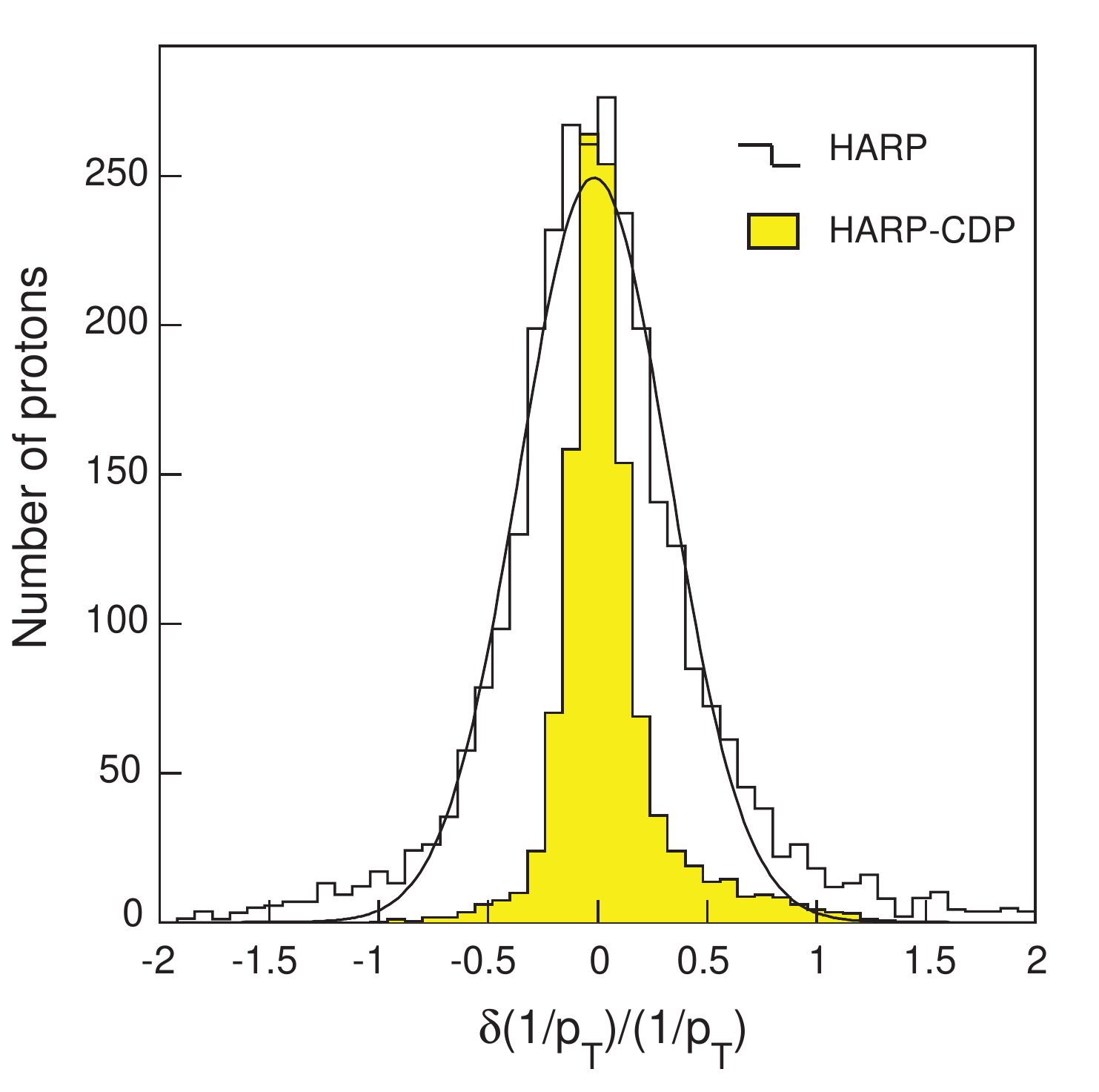}
\caption{Resolutions $\delta (1/p_{\rm T})/(1/p_{\rm T})$ from
HARP--CDP (shaded histogram) and from the HARP Collaboration (open
histogram).} 
\label{resolutions}
\end{center}
\end{figure}

We show in figure~\ref{resolutions} with 
the shaded histogram the absence of any momentum bias, and
the momentum resolution, obtained by our group in the 
elastic scattering of 3~GeV/{\it c} pions and protons on 
protons at rest. Our resolution, from fits with the beam point included, 
is $\sigma (1/p_{\rm T}) \sim 0.20$~(GeV/{\it c})$^{-1}$, 
well consistent with what is
expected from our TPC calibration work~\cite{TPCpub}.    
It is unclear why the authors avoid proving their claim 
of a resolution of $\sigma (1/p_{\rm T} \sim 0.30$~(GeV/{\it c})$^{-1}$ 
by showing their analogous distribution. Rather, they argue 
their case with the much worse resolution from fits without 
the beam point (although the authors' missing-mass analysis is based on
fits with the beam point). For comparison, their data (copied from the 
middle panel of figure~6 in Ref.~\cite{arXivPreprintTPC}), 
are shown as open histogram in figure~\ref{resolutions}.
Superimposed on their data is a Gaussian fit with 
$\sigma = 0.33$. With an approximate $p_{\rm T} = 0.45$~GeV/{\it c}
the authors' resolution is 
$\sigma (1/p_{\rm T}) = 0.73$~(GeV/{\it c})$^{-1}$, 
worse than the 0.60~(GeV/{\it c})$^{-1}$ expected 
for fits without beam points. This is consistent with the
evidence shown in Section~\ref{pTdiscussion} that their resolution 
$\sigma (1/p_{\rm T})$ is much worse than 
0.30~(GeV/{\it c})$^{-1}$. 

\section{Concluding commentary}

We presented evidence of serious defects in the  
large-angle data analysis of the HARP Collaboration: 
(i) the $p_{\rm T}$ scale 
is systematically biased by 
$\Delta(1/p_{\rm T}) \sim 0.3$~(GeV/{\it c})$^{-1}$;
(ii) the $p_{\rm T}$ resolution is by a 
factor of two worse than claimed; and
(iii) the discovery of the `500~ps effect' 
in the HARP multi-gap RPCs is false. 

In defiance of explicit and repeated criticism 
of their work at various levels, including published 
`Comments'~\cite{NIMComments}--\cite{CommentsOnTaPaper},
HARP keep insisting on the validity of their 
work~\cite{NIMRebuttal,IEEERebuttal}. 

Yet HARP have been unable to disprove any of the critical arguments against their results.
Their arguments in their defence confirm, rather than disprove, our claims of serious defects in their large-angle data analysis.

In this unusual and regrettable situation, we  
warn the community that cross-sections that are based on the
TPC and RPC calibrations reported by HARP, are wrong
by factors of up to two.

\end{document}